\documentclass{easychair}

\usepackage{makeidx}

\usepackage[numbers]{natbib}
\usepackage{bibentry}
\nobibliography*

\newcommand{\xf}[1]{Figure~\ref{#1}}

\newcommand{\xs}[1]{Section~\ref{#1}}
\newcommand{\xa}[1]{Appendix~\ref{#1}}

\newcommand{\xt}[1]{Table~\ref{#1}}
\newcommand{\xl}[1]{Listing~\ref{#1}}

\newcommand{\AST}{{AST\index{AST}}}

\newcommand{\gipc}{{GIPC\index{GIPC}\index{Frameworks!GIPC}}}
\newcommand{\gicf}{{GICF\index{GICF}\index{Frameworks!GICF}}}

\newcommand{\gee}{{GEE\index{GEE}\index{Frameworks!GEE}}}

\newcommand{\gipsy}{{GIPSY\index{GIPSY}}}

\newcommand{\glu}{{GLU\index{GLU}}}
\newcommand{\glusharp}{{GLU\#\index{GLU\#}}}
\newcommand{\gipl}{{GIPL\index{GIPL}}}

\newcommand{\lucid}{{Lucid\index{Lucid}}}
\newcommand{\ilucid}{{Indexical Lucid\index{Indexical Lucid}}}
\newcommand{\jlucid}{{JLucid\index{JLucid}}}
\newcommand{\olucid}{{Objective Lucid\index{Tensor Lucid}}}
\newcommand{\tlucid}{{Tensor Lucid\index{Tensor Lucid}}}
\newcommand{\plucid}{{Partial Lucid\index{Partial Lucid}}}
\newcommand{\flucid}{{Forensic Lucid\index{Forensic Lucid}}}
\newcommand{\onyx}{{Onyx\index{Onyx}}}
\newcommand{\lucx}{{Lucx\index{Lucx}}}

\newcommand{\jooip}{{JOOIP\index{JOOIP}}}
\newcommand{\marfl}{MARFL\index{MARFL}}
\newcommand{\translucid}{TransLucid\index{TransLucid}}
\newcommand{\lustre}{Lustre\index{Lustre}}

\newcommand{\C}{{C\index{C}}}
\newcommand{\cpp}{{C++\index{C++}}}
\newcommand{\perl}{{Perl\index{Perl}}}
\newcommand{\java}{{Java\index{Java}}}
\newcommand{\python}{{Python\index{Python}}}
\newcommand{\fortran}{{Fortran\index{Fortran}}}

\newcommand{\todo}[0]
{
	{\Large \[TODO\]}
}

\newcommand{\tool}[1]{\texttt{#1}\index{Tools!#1}}

\newcommand{\api}[1]{\texttt{#1}\index{API!#1}}

\newcommand{\codesegment}[1]{\texttt{\##1}\index{Segments!\##1}}

\newcommand{\lucidL}[1]{{$\mathit{Lucid}$}($L$) }

		{}

\def\myvert{\raise 2.27pt \hbox{\vrule depth 0pt height 8pt width 0.2mm}}
\def\myarrow{\hspace*{0.43mm}%
             \raise 2.29pt\hbox{\vrule depth 0pt height 8pt width 0.16mm}%
             \hspace*{-0.32mm}%
             $\longrightarrow$
             \ %
             }

\lstdefinestyle{codeStyle}
{
	language=Java,
	frame=single,  
	basicstyle=\footnotesize,
	captionpos=b,
	showstringspaces=false,
	showspaces=false,
	extendedchars=true,
	linewidth=1\linewidth,
	breaklines=true,
	float=phtb  
}

\makeindex

\begin{document}

%
%


\title{Furthering Baseline Core Lucid Standard Specification
       in the Context of the History of Lucid, Intensional Programming,
       and Context-Aware Computing}
\titlerunning{Core Lucid Standard Specification: SIGLUCID's Contextual Design}

{\author
	{
		Paquet, Joey\\
		Mokhov, Serguei A.\\\\SIGLUCID
	}
}
\authorrunning{SIGLUCID}

\maketitle

%
%

\begin{abstract}
This work is multifold. We review the historical literature on the {\lucid}
programming language, its dialects, intensional logic, intensional programming,
the implementing systems, and context-oriented and context-aware computing
and so on that provide a contextual framework for the converging Core Lucid
standard programming model.
We are designing a standard specification of a baseline {\lucid} virtual machine
for generic execution of Lucid programs. The resulting Core Lucid language
would inherit the properties of generalization attempts of {\gipl} (1999--2013)
and {\translucid} (2008--2013) for all future and recent Lucid-implementing
systems to follow.
We also maintain this work across local research group in order to
foster deeper collaboration, maintain a list of recent and historical bibliography
and a reference manual and reading list for students.
We form a (for now informal) SIGLUCID group to keep track of this standard and historical records
with eventual long-term goal through iterative revisions for this work to become
a book or an encyclopedia of the referenced topics, and perhaps, an RFC.
We first begin small with this initial set of notes.
\end{abstract}

\tableofcontents
\listoffigures
\lstlistoflistings
\listoftables

\section{Introduction}
\label{sect:introduction}

This work gears toward a generalization on a number of previous results by various authors
in terms of context specification in the {\lucid} programming language and the Core Lucid
dealing with data types and have a virtual machine standard agreed to by SIGLUCID.
Aside from various data types (primarily to address the hybrid computing paradigms
uniformly of Lucid integrating with imperative dialects), the context definition should
also be hierarchical for certain application domains to allow for context nesting.
The notion of context is central to {\lucid} as an explicit meaning component
that is specified as a first class-value. Traditional {\lucid}'s context
specification was assuming tags and the corresponding values were
{\em simple} -- i.e. a collection of dimension names and the value pairs
would denote a point in the context space. Then, the notion of point
was not sufficient for some Lucid dialects that needed higher-order
contextual notions, such as context sets to denote a context area
or field instead of a point, as it was done in {\lucx}. Another way
to traverse a more complex notion of the context definition was
done in iHTML and related tools where nesting of the tags would
denote the nesting of contextual expressions forming a sort of
contextual tree, where the actual tag values were at the leaves
of the tree.  Then, a similar need arose in {\flucid} and {\marfl}
to specify higher-order contexts representing evidence and witness
stories or configuration details, but allowing evaluation at any level
of the context tree rather than just the leaves. Thus, this work aims at
unifying and standardizing various context specifications under one
uniform intermediate form that all Lucid dialects can adhere to
thereby making the community speak the same language and
potentially bring interoperability between various Lucid implementations
and incarnations across University groups working in the
intensional programming domain.

\subsection{Motivation}

Higher-order context specification is needed for nested-level
context that traditionally decomposes a higher-order value
into its components, and equivalently from the components
get to the parent component. This is partitioned in any
nested markup-like language, e.g. iHTML, any XML-based
definitions and descriptions of data and databases,
configuration management of a software system components,
as well as domain-specific applications such as contextual
specification of a cyberforensic case where evidential statement
is comprised of observation sequences representing encoded stories
told by evidence and witnesses, which in turn decompose into
observations, and then into properties and duration components;
which all-in-all comprise a context of evaluation of a cyberforensic
case. Thus, the need for higher-order contexts is apparent as
a fundamental pillar supporting higher-order intensional
logic (HOIL).

Types other than the context also should be exposed to the
programmer when needed and allow for a wider range of data
types and type systems to allow hybrid dialect interaction
easier as well as compiler optimization and run-time system
parallelizations.


\subsection{Proposed Solution}

For the context specification, we propose to extend the notion of context to be a bi-directional
tree with the operators from {\gipl}, {\lucx}, iHTML and {\marfl} to query,
switch, and traverse the depth of the context hierarchy. The language
that encompasses the new specification on the syntax and semantic
level is proposed to be called {\em Core Lucid} or {\em Standard Lucid} or {\em Nominal Lucid}. 

The type specification and code segments are augmented as presented in possible
specification from the SIGLUCID meetings and others in \xs{sect:siglucid-meetings}
and \xs{sect:gipsy-hybrid-interaction}.


\subsection{SIGLUCID}
\label{sect:siglucid}

SIGLUCID: Special Interest Group on Lucid, Ubiquity,
Context, Intensionality, and multi-Dimensionality.
SIGLUCID is a working group of researchers in Lucid,
intensional programming, intensional logic, context-aware
and context-oriented computing and the related application
domains (see \xf{sect:applications}.

\noindent
SIGLUCID currently is a loose affiliation of researchers,
collaborators, and supporters in intensional logic, intensional
programming, context-aware computing, etc. across Canada,
Australia, and other places.

Should you wish to be a part and contribute, contact the people listed
at the title page. This is a running draft to fill in the missing
information as it becomes possible.

%
%
%

\section{Historical Perspective, Context, Dialects, and Applications}
\label{sect:history}

The history of {\lucid}, multi-dimensional intensional programming and logic,
context-orientation, parallel, concurrent, and distributed eductive
evaluation aspects can be traced through different Lucid dialects,
outlined in \xs{sect:lucid-dialects}.

\subsection{Lucid Dialects}
\label{sect:lucid-dialects}

Here we enumerate the Lucid dialects that came to be from either
practical implementations and/or theoretical frameworks to study
the intensionality properties, context, and mathematical and
intensional logic foundations. We plan to make the list into
a table or other presentation means with the status of each language
and the related citations.

\begin{itemize}
	\item {\lucid}
	\item {\gipl}
	\item {\translucid}
	\item {\lucx}
	\item {\glu}
	\item {\glusharp}
	\item {\ilucid}
	\item {\tlucid}
	\item {\plucid}
	\item {\jlucid}
	\item {\olucid}
	\item {\onyx}
	\item {\flucid}
	\item {\jooip}
	\item {\marfl}
	\item IHTML
	\item IHTML2
	\item iPerl
	\item ISE
	\item vmake
	\item {\lustre}
	\item pLucid
	\item Luthid
\end{itemize}

\subsubsection{Incomplete Brief History and The Family}
\index{Lucid!History}
\index{Lucid!Family}

From 1974 to {\lucid} Today (taken from \cite{mokhovmcthesis05}, incomplete,
to be updated:

\begin{enumerate}
\item
Lucid as a Pipelined Dataflow Language\index{Lucid!Pipelined Dataflows} through 1974-1977.
{\lucid} was introduced by Anchroft and Wadge in \cite{lucid76,lucid77}. Features:
	\begin{itemize}
	\item
	A purely declarative language for natural expression of iterative algorithms.

	\item
	Goals: semantics and verification of correctness of programming languages
	(for details see \cite{lucid76,lucid77}).

	\item
	Operators as pipelined streams: one for initial element, and then all for the
	successor ones.
	\end{itemize}

\item
Intensions, {\ilucid}, GRanular Lucid ({\glu}, \cite{glu1,glu2}), circa 1996.
More details on these two dialects are provided further in the chapter as
they directly relate to the theme of this thesis. Features:
	\begin{itemize}
	\item
	Random access to streams in {\ilucid}.

	\item
	First working hybrid intensional-imperative	paradigm
	({\C}/{\fortran} and {\ilucid}) in the form of {\glu}.

	\item
	Eduction or demand-driven execution (in {\glu}).
	\end{itemize}

\item
{\plucid}, {\tlucid}, 1999 \cite{paquetThesis}.
	\begin{itemize}
	\item
	{\plucid}
	is an intermediate experimental language used for
	demonstrative purposes in presenting the semantics of
	{\lucid} in \cite{paquetThesis}.

	\item
	{\tlucid} dialect was developed by Joey Paquet for plasma
	physics computations to illustrate advantages and expressiveness
	of {\lucid} over an equivalent solution written in {\fortran}.
	\end{itemize}

\item
{\gipl}, 1999 \cite{paquetThesis}.
	\begin{itemize}
	\item
		All Lucid dialects can be translated into
		this basic form of {\lucid}, {\gipl} through
		a set of translation rules. ({\gipl} is in the
		foundation of the execution semantics
		of {\gipsy} and its {\gipc} and {\gee} because its
		{\AST} is the only type of {\AST} {\gee} understands
		when executing a GIPSY program).
	\end{itemize}

\item RLucid, 1999, \cite{rlucid99}
	\begin{itemize}
	\item
		A Lucid dialect for reactive real-time
		intensional programming.
	\end{itemize}

\item
{\jlucid}, {\olucid}, 2003 - 2005
	\begin{itemize}
	\item
		These dialects introduce a notion of hybrid and object-oriented programming
		in the {\gipsy} with {\java} and {\ilucid} and {\gipl},
		and are discussed great detail in the follow up chapters
		of this thesis.
	\end{itemize}


\item
{\lucx} \cite{kaiyulucx}, 2003 - 2005
	\begin{itemize}
	\item
		Kaiyu Wan introduces a notion of contexts as first-class values
		in {\lucid}, thereby making {\lucx} the true intensional language.
	\end{itemize}

\item
{\onyx} \cite{grogonoonyx2004}, April 2004.
	\begin{itemize}
	\item
		Peter Grogono makes an experimental derivative of {\lucid} -- {\onyx}
		to investigate on lazy evaluation of arrays.
	\end{itemize}

\item
{\glusharp} \cite{glu3}, 2004
	\begin{itemize}
	\item
		{\glusharp} is an evolution of {\glu} where {\lucid}
		is embedded into {\cpp}.
	\end{itemize}
\end{enumerate}

\subsection{List of Tools and Implementing Systems}

\begin{enumerate}
	\item {\gipsy} \cite{gipsy-all-named}
	\item {\glu}
	\item {\translucid}
	\item pLucid 
	\item \tool{libintense}
\end{enumerate}

\subsection{Application Domains}
\label{sect:applications}

\begin{enumerate}
	\item Context-Aware Computing
	\item Scientific Computing
	\item Distributed and Parallel Evaluation
	\item Ubiquitous and Mobile Computing
	\item Wiki
	\item Forensic Computing
	\item Multimedia and Configuration Management
	\item Program Verification
	\item Software Engineering
	\item Aspect-Oriented Programming
	\item Web OS
	\item Reactive Computing
	\item Pervasive Computing
	\item Autonomic Computing
	\item Modeling and Simulation
	\item Model Checking
\end{enumerate}

\subsection{Related Work}
\label{sect:related-work}

There is a vast amount of related and past work done. Over time we will provide
brief historical description of each or a group of works clustered by a specific
theme either in this section or relevant other sections.
For now, however, we begin by citing them first, so anyone looking for the references
can look them up in a jiffy and make their choice accordingly. This is ideal for
graduate students and researchers starting in the subjects or looking for what's
been done that they can benefit from.

Most recent on top:

\begin{itemize}

\item 2013
\begin{itemize}
	\item
\bibentry{mokhov-phd-thesis-2013}
	\item
\bibentry{graph-based-gmt}
\end{itemize}

\item 2012
\begin{itemize}
	\item
\bibentry{unifying-refactoring-jini-jms-dms}
\end{itemize}

\item 2011
\begin{itemize}
	\item
\bibentry{plaice-habilitation-2011}
	\item
\bibentry{ji-yi-mcthesis-2011}
	\item
\bibentry{flucid-printer-case-icdf2c-2011}
	\item
\bibentry{flucid-dfg-viz-pst2011}
\end{itemize}

\item 2010
\begin{itemize}
	\item
\bibentry{gipsy-type-system-hoil-theory}
	\item
\bibentry{gipsy-hoil}
	\item
\bibentry{gipsy-jooip}
	\item
\bibentry{gipsy-multi-tier-impl}
	\item
\bibentry{bin-han-10}
	\item
\bibentry{unifying-refactoring-jini-jms-dms}
	\item
\bibentry{mokhov-mcthesis-book-reprint10}
	\item
\bibentry{flucid-raid2010}
	\item
\bibentry{flucid-dfg-viz}
	\item
\bibentry{self-forensics-jooip-flucid-assl}
\end{itemize}

\item 2009
\begin{itemize}
	\item
\bibentry{gipsy-multi-tier-secasa09}
	\item
\bibentry{gipsy-type-system-c3s2e09}
	\item
\bibentry{aihuawu09}
	\item
\bibentry{flucid-blackmail-hsc09}
	\item
\bibentry{flucid-printer-case-hsc09}
	\item
\bibentry{self-forensics-through-case-studies}
	\item
\bibentry{self-forensics-flucid-road-vehicles}
	\item
\bibentry{flucid-credibility-wips}
	\item
\bibentry{self-forensics-nasa-rfi}
	\item
\bibentry{auto-synth-deploy-intensional-kahn-nets-2009}
\end{itemize}

\item 2008
\begin{itemize}
	\item
\bibentry{ben-hamed-phd-08}
	\item
\bibentry{translucid-cartesian-ips-2008}
	\item
\bibentry{knowl-rep-temporal-logic-ips-2008}
	\item
\bibentry{lucid-dataflow-oo-ips-2008}
	\item
\bibentry{possible-world-versioning-ips-2008}
	\item
\bibentry{efficient-intensional-lazy-eval-func-langs-ips-2008}
	\item
\bibentry{pourteymourmcthesis08}
	\item 
\bibentry{tongxinmcthesis08}
	\item 
\bibentry{marfl-context-secasa08}
	\item 
\bibentry{gipsy-simple-context-calculus-08}
	\item 
\bibentry{eager-translucid-secasa08}
	\item 
\bibentry{multithreaded-translucid-secasa08}
	\item 
\bibentry{dms-pdpta08}
	\item 
\bibentry{agipsy-ease08}
	\item 
\bibentry{flucid-isabelle-techrep-tphols08}
	\item 
\bibentry{flucid-imf08}
	\item 
\bibentry{marf-into-flucid-cisse08}
	\item 
\bibentry{marf-gipsy-distributed-ispdc08}
\end{itemize}

\item 2007
\begin{itemize}
	\item 
\bibentry{dmf-pdpta07}
	\item 
\bibentry{gipsy-context-calculus-07}
	\item
\bibentry{ditu-translucid-2007}
\end{itemize}

\item 2006
\begin{itemize}
	\item 
\bibentry{wanphd06}
\end{itemize}

\item 2005
\begin{itemize}
	\item
\bibentry{real-time-reactive-prog-lucid-2005}
	\item 
\bibentry{kaiyulucx}
	\item 
\bibentry{mokhovgicf2005}
	\item 
\bibentry{mokhovmcthesis05}
	\item 
\bibentry{mokhovolucid2005}
	\item 
\bibentry{mokhovjlucid2005}
	\item 
\bibentry{gipsy2005}
	\item 
\bibentry{dmf-cnsr05}
	\item 
\bibentry{wu05}
	\item 
\bibentry{dmf-plc05}
	\item 
\bibentry{vassev-mscthesis-05}
	\item 
\bibentry{crr05}
	\item 
\bibentry{foil-axiomatized-2005}
\end{itemize}

\item 2004
\begin{itemize}
	\item 
\bibentry{swobodaphd04}
	\item 
\bibentry{bolu04}
	\item 
\bibentry{distributed-context-computing}
	\item 
\bibentry{active-functional-idatabase}
	\item 
\bibentry{grogonoonyx2004}
	\item 
\bibentry{wuf04}
	\item 
\bibentry{leitao04}
	\item 
\bibentry{yimin04}
	\item 
\bibentry{glu3}
	\item 
\bibentry{ip-for-agent-communication-dalt2004}
\end{itemize}

\item 2003
\begin{itemize}
	\item 
\bibentry{bolu03}
	\item 
\bibentry{wu03}
	\item 
\bibentry{wadgeHammings}
	\item 
\bibentry{middleware-context-aware-agents-2003}
	\item 
\bibentry{intensional-extensional-semantics-dataflow-2003}
\end{itemize}

\item 2002
\begin{itemize}
	\item 
\bibentry{gipcincrements}
	\item 
\bibentry{aihuawu02}
	\item 
\bibentry{chunleiren02}
\end{itemize}

\item 2000
\begin{itemize}
	\item 
\bibentry{gipsy-arch-2000}
\end{itemize}

\item 1999
\begin{itemize}
	\item 
\bibentry{kropf-plaice-intensional-objects-99}
	\item 
\bibentry{yotis99}
	\item 
\bibentry{paquetThesis}
	\item
\bibentry{intensional-programming-2}
	\item
\bibentry{intensional-logic-context-1999}
	\item
\bibentry{constraint-solving-rules-1999}
	\item
\bibentry{hyper-hypertext-1999}
	\item
\bibentry{intensional-hypertext-1999}
	\item
\bibentry{associative-query-web-1999}
	\item
\bibentry{web-server-groups-dygop-1999}
	\item
\bibentry{two-semantics-temporal-constraint-logic-1999}
	\item
\bibentry{agent-grouping-executable-temporal-logic-1999}
	\item
\bibentry{negation-linear-time-temporal-logic-1999}
	\item
\bibentry{branching-time-logic-cactus-1999}
	\item
\bibentry{temporal-reasoning-trl-1999}
	\item
\bibentry{multi-formalism-distrib-dataflow-context-1999}
	\item 
\bibentry{rlucid99}
	\item
\bibentry{mining-time-series-db-1999}
	\item
\bibentry{temporal-meaning-representation-nlp-1999}
	\item
\bibentry{statistical-queries-historical-db-1999}
	\item
\bibentry{periodic-phenomena-exceptions-1999}
	\item
\bibentry{intensional-model-large-scale-distrib-sys-1999}
	\item
\bibentry{semantics-dimensions-values-1999}
	\item
\bibentry{adding-multidimensionality-to-pls-1999}
	\item
\bibentry{intensional-communities-1999}
	\item 
\bibentry{intensionalisation-tools}
\end{itemize}

\item 1998
\begin{itemize}
	\item 
\bibentry{ihtml-1998}
\end{itemize}

\item 1997
\begin{itemize}
	\item 
\bibentry{zhao-ooip-97}
	\item 
\bibentry{glu2}
	\item 
\bibentry{multi-dim-logic-programming-theory-1997}
\end{itemize}

\item 1996
\begin{itemize}
	\item 
\bibentry{dodd96}
	\item 
\bibentry{glu1}
\end{itemize}

\item 1995
\begin{itemize}
	\item
\bibentry{paquet-intensional-databases-95}
	\item
\bibentry{lucid95}
	\item
\bibentry{intensional-programming-1}
	\item
\bibentry{functional-ext-lustre-1995}
	\item
\bibentry{multidim-prog-verification-1995}
	\item
\bibentry{relative-debugging-multi-prog-versions-1995}
	\item
\bibentry{possible-woorlds-1995}
	\item
\bibentry{glu-graphical-models-1995}
	\item
\bibentry{prog-distrib-systems-based-on-graphs-1995}
	\item
\bibentry{logic-into-prog-langs-1995}
	\item
\bibentry{alfa-dataflow-machine-1995}
	\item
\bibentry{agi95glu}
	\item
\bibentry{particle-in-cell-sim-lucid-1995}
	\item
\bibentry{realtime-oo-spec-verification-1995}
	\item
\bibentry{verify-multran-progs-temporal-logic-1995}
	\item
\bibentry{possible-www-1995}
	\item
\bibentry{intensional-relation-1995}
	\item
\bibentry{intensional-algo-higher-order-progs-1995}
	\item
\bibentry{async-calculus-absence-actions-1995}
	\item
\bibentry{choice-first-class-1995}
	\item
\bibentry{meta-level-modal-logic-programming-1995}
	\item
\bibentry{fuzzy-temporal-prolog-1995}
	\item
\bibentry{knowledge-based-sim-chronolog-1995}
\end{itemize}

\item 1994
\begin{itemize}
	\item
\bibentry{yotis94}
	\item
\bibentry{indexicalQuery-94}
	\item
\bibentry{sTaoThesis}
	\item
\bibentry{du-oo-ipl-impl-94}
\end{itemize}

\item 1993
\begin{itemize}
	\item 
\bibentry{new-app-version-plaice-93}
	\item 
\bibentry{absToRealTime-Plaice-93}
	\item 
\bibentry{multi-dim-problem-solving-lucid-1993}
\end{itemize}

\item 1991
\begin{itemize}
	\item 
\bibentry{duThesis}
	\item 
\bibentry{freeman-benson-lobjcid-91}
\end{itemize}

\item 1990
\begin{itemize}
	\item 
\bibentry{eductive-3dspreadsheet-Du-90}
	\item 
\bibentry{3dspreadsheet-Du-90}
\end{itemize}

\item 1989
\begin{itemize}
	\item 
\bibentry{Real-Time-Dataflow}
\end{itemize}

\item 1988
\begin{itemize}
	\item 
\bibentry{manual-il-88}
\end{itemize}

\item 1987
\begin{itemize}
	\item 
\bibentry{nasa-autogen-lucid87}
	\item 
\bibentry{eductive-interpreter}
\end{itemize}

\item 1985
\begin{itemize}
	\item 
\bibentry{lucid85}
\end{itemize}

\item 1982
\begin{itemize}
	\item
\bibentry{denotational-operational-semantics-dataflow}
	\item
\bibentry{r-for-semantics-82}
\end{itemize}

\item 1981
\begin{itemize}
	\item 
\bibentry{luthid}
\end{itemize}

\item 1977
\begin{itemize}
	\item 
\bibentry{nonprocedural-iterative-lucid-77}
	\item 
\bibentry{lucid77}
\end{itemize}

\item 1976
\begin{itemize}
	\item 
\bibentry{lucid76}
\end{itemize}

\end{itemize}


\noindent
Wikipedia and other Wiki entries:

\begin{itemize}
	\item \url{http://en.wikipedia.org/wiki/Lucid_(programming_language)}
	\item \url{http://en.wikipedia.org/wiki/Category:Intensional_programming_languages}
	\item \url{http://en.wikipedia.org/wiki/Intensional_logic}
	\item \url{http://www.haskell.org/haskellwiki/Lucid}
\end{itemize}

\section{Core Lucid Standard Specification Design}

The Core Lucid standard design and specification is an ongoing process
influenced by the two core proposals: {\gipl} and {\translucid}
developed in 1999 and 2008 respectively.

\subsection{SIGLUCID Meetings}
\label{sect:siglucid-meetings}

Here's the brief summary of the SIGLUCID meetings at various workshops
and conferences, attendees, and works contributing to the collaboration
and developing the Core Lucid standard.

\subsubsection{SECASA 2010 Meeting at SERA 2010, Montreal, Canada}
\label{sect:meeting-notes-secasa2010}

\paragraph{Works}

The following works were presented:

\begin{enumerate}
	\item \cite{gipsy-multi-tier-impl}
	\item \cite{gipsy-type-system-hoil-theory}
	\item \cite{gipsy-jooip}
	\item \cite{gipsy-hoil}
\end{enumerate}

{\todo}

\paragraph{Attendees}

\begin{enumerate}
	\item Serguei A. Mokhov
	\item Joey Paquet
	\item Emil Vassev
	\item Bin Han
\end{enumerate}

{\todo}

\subsubsection{SECASA 2009 Meeting at COMPSAC 2009, Seattle, USA}
\label{sect:meeting-notes-secasa2009}

\paragraph{Works}

The following works were presented:

\begin{enumerate}
	\item \cite{gipsy-multi-tier-secasa09}
\end{enumerate}

{\todo}

\paragraph{Attendees}

\begin{enumerate}
	\item Joey Paquet
	\item John Plaice
\end{enumerate}

{\todo}

\subsubsection{SECASA 2008 Meeting at COMPSAC 2008, Turku, Finland}
\label{sect:meeting-notes-secasa2008}

The first discussion about standardizing the types,
evaluation, and overview of the current candidates
for the Lucid Core from different research groups,
such as {\gipl}, {\translucid}. The needs of
various in-progress Lucid dialects were discussed
to be accommodated in the core, such as {\marfl},
{\flucid}. Below are the points from the meeting minutes.

\paragraph{Works}

The following works were presented:

\begin{enumerate}
	\item \cite{gipsy-simple-context-calculus-08}
	\item \cite{eager-translucid-secasa08}
	\item \cite{multithreaded-translucid-secasa08}
	\item \cite{marfl-context-secasa08}
\end{enumerate}

\paragraph{Attendees}

\begin{enumerate}
	\item Weichang Du
	\item Blanca Mancilla
	\item Serguei A. Mokhov
	\item Joey Paquet
	\item John Plaice
	\item Toby Rahilly
	\item William W. Wadge
\end{enumerate}

\paragraph{Notes}

\begin{enumerate}

\item
Constants appearing in the expressions:

\begin{verbatim}
type<string> (const type -- int8<42> != int16<42>)
[| string |]
12
true
false
\end{verbatim}

\item
header -- default types for integers, etc.

\item
\#parens -- see the sections on the GIPSY
type system and a hybrid program example \xs{sect:gipc-preprocessor} and \xs{sect:gipsy-types}

\item
Proposal of \api{type} ('\api{type}' is a keyword).

\begin{itemize}
\item
type$<$float 32$>$
\item
uchar$<${{{$>$ uchar$<$}}}$>$
\item shorthand syntactical sugar:

\verb+[||| 1.2 |||]+ -- 64 bit IEEE float

\verb+{{{ 1.2 }}}+ -- 32 bit float
\end{itemize}

\item
\verb+special<...>+ -- correspond to exceptions and
error situations for handling later on

\item
Joey: Dimensions syntactically are allowed to taken on default values
other than always implicit default of zero.

\begin{itemize}
\item
\verb+special<undecl> + special<arith> = special<undecl>+

 -- lose details, e.g. where it happened in the code
 or even within the imperative code?

\item
\begin{verbatim}
if(isspecial<undecl> E)
then ...
\end{verbatim}
\end{itemize}

\item
Bill has done something like that with someone in pLucid, with
well defined semantics, etc.

\item
Bill complained about eagerness:

$$\{E_1:E_2, ..., E_n:E_{n_{2}}\}$$

eager: only lefts (multithreaded~\cite{multithreaded-translucid-secasa08}),
else all ---, but right-hand-sides are lazy.

\item
Q: How to stop people from producing recursive/infinite contexts?

\item
Bill: risky: \verb+\# a@{a:P, E:Q} != P+ ???
If $E$ does not terminate or \api{special} -- can't prove it's constant.

\item
Toby: threading, sequential scheduling 

\item Audience concluded:
{\gipl} context-eager,
{\translucid} dimension-eager (LHS)

\item
Variables

\begin{verbatim}
variable x
dimension d
\end{verbatim}

type of $x$ is context-dependent.

Future:

id$<$x$>$

dimension$<$d$>$

expr$<$E$>$

\item
Dynamic ranks analysis
(Tony Faustini and Bill in the '80)

$$(X, C)?$$

$? =$ eduction evaluation engine

\begin{verbatim}
W?(X, {})
 ->42
 ->{d1,....,dn}?
\end{verbatim}

\begin{verbatim}
v1=C(d1)
vn=C{dn}
v'1=C(d1')
v'm=C{dm'}
\end{verbatim}

\begin{verbatim}
W?(X, {d1:v1, ....})
 ->42
 -> {d'2,...,d'n}?
\end{verbatim}

\begin{verbatim}
x, W, C, Cs, Ci # 3
W' = W U {(X, Cs) |-> {3}?}
\end{verbatim}

$C_i$ -- current internal context

\item
Toby: optimization: demand grouping
demands as early as possible, lifting up

\item
optimization for constant vs. run-time dimensions
thus

\texttt{dimension d}

is an optimization hint.

\item
Binary representation (portable) ???
a-la Java byte-code
\end{enumerate}

\subsubsection{PLC 2005 Meeting at WORLDCOMP 2005, Las Vegas, USA}
\label{sect:meeting-notes-plc2005}

\paragraph{Works}

The following works were presented:

\begin{enumerate}
	\item \cite{mokhovjlucid2005}
	\item \cite{mokhovolucid2005}
	\item \cite{mokhovgicf2005}
	\item \cite{dmf-plc05}
	\item \cite{wu05}
	\item \cite{gipsy2005}
	\item \cite{kaiyulucx}
\end{enumerate}

{\todo}

\paragraph{Attendees}

\begin{enumerate}
	\item Weichang Du
	\item Serguei A. Mokhov
	\item Joey Paquet
	\item Emil Vassev
	\item William W. Wadge
	\item Kaiyu Wan
	\item Aihua Wu
\end{enumerate}

{\todo}

\subsubsection{ISLIP 1999}
\label{sect:meeting-notes-islip1999}

\paragraph{Works}

The following works were presented:

\begin{enumerate}
	\item \cite{intensional-programming-2}
\end{enumerate}

{\todo}

\paragraph{Attendees}

\begin{enumerate}
	\item William W. Wadge
	\item John Plaice
	\item Joey Paquet
	\item ...
\end{enumerate}

{\todo}

\subsubsection{ISLIP 1995}
\label{sect:meeting-notes-islip1995}

\paragraph{Works}

The following works were presented:

\begin{enumerate}
	\item \cite{intensional-programming-1}
\end{enumerate}

{\todo}

\paragraph{Attendees}

\begin{enumerate}
	\item William W. Wadge
	\item John Plaice
	\item Joey Paquet
	\item ...
\end{enumerate}

{\todo}

\subsection{{\translucid}}
\label{sect:translucid-reqs-for-standard}

{\todo}

\subsection{{\gipsy}}
\label{sect:gipsy-reqs-for-standard}

\subsubsection{Hybrid Interaction with Other Languages}
\label{sect:gipsy-hybrid-interaction}

\paragraph{GIPC Preprocessor}
\label{sect:gipc-preprocessor}
\index{Preprocessor}
\index{GIPC!Preprocessor}
\index{Preprocessor!GIPC}

\begin{figure}
	\begin{centering}
	\includegraphics[width=\textwidth]{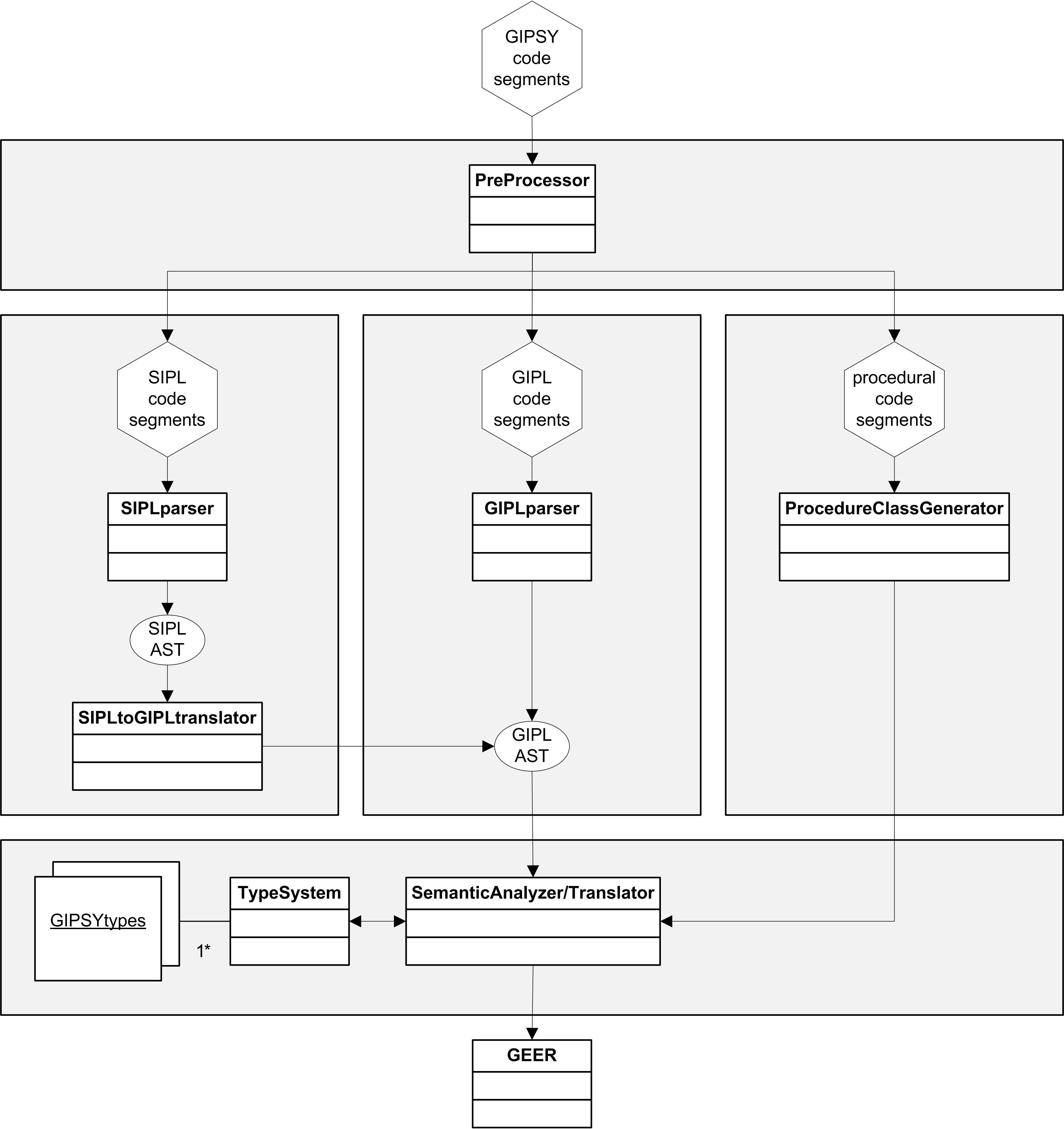}
	\caption{GIPC Framework with the Preprocessor}
	\label{fig:gipc-preprocessor}
	\end{centering}
\end{figure}

\begin{figure}
	\begin{centering}
	\includegraphics[totalheight=.9\textheight]{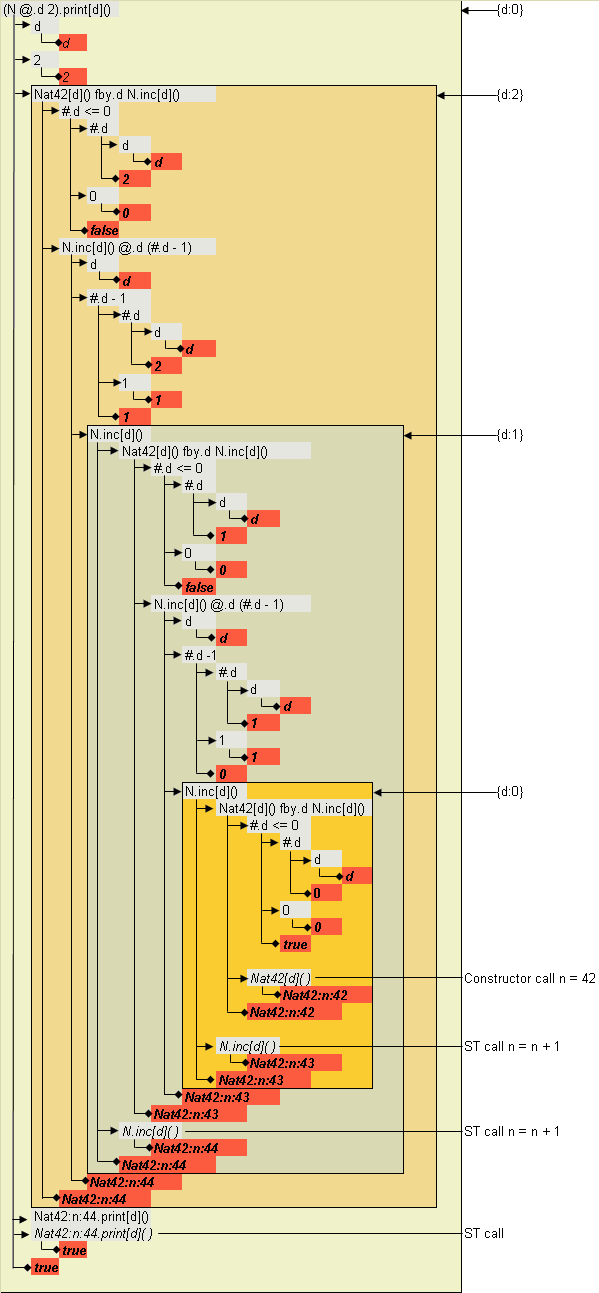}
	\caption{Example of Eductive Evaluation of {\olucid} Progran}
	\label{fig:exegraphobj}
	\end{centering}
\end{figure}

The \api{Preprocessor} \cite{mokhovmcthesis05,mokhovgicf2005} is something that is invoked first by the GIPC (see \xf{fig:gipc-preprocessor}) on incoming GIPSY program's source code stream. The \api{Preprocessor}'s role is to do preliminary program analysis, processing, and splitting the source GIPSY program into ``chunks'', each written in a different language and identified by a {\it language tag}. In a very general view, a GIPSY program is a hybrid program consisting of different languages in one or more source file; then, there has to be an interface between all these code segments. Thus, the \api{Preprocessor} after some initial parsing (using its own preprocessor syntax) and producing the initial parse tree, constructs a preliminary dictionary of symbols used throughout the program. This is the basis for type matching and semantic analysis applied later on. This is also where the first step of type assignment occurs, especially on the boundary between typed and typeless parts of the program, e.g. {\java} and a specific Lucid dialect. The \api{Preprocessor} then splits the code segments of the GIPSY program into chunks preparing them to be fed to the respective concrete compilers for those chunks. The chunks are represented through the \api{CodeSegment} class that the \api{GIPC} collects.

\paragraph{GIPSY Program Segments}
\index{GIPSY Program!Segments}

There are four baseline types of segments defined to be used in a GIPSY program. These are:

\begin{itemize}
\item \codesegment{funcdecl}
program segment declares function prototypes written as imperative language functions defined later or externally from this program to be used by the intensional language part. The syntactical form of these prototypes is particular to GIPSY programs and need not resemble the actual function prototype declaration they describe in their particular programming language. They serve as a basis for static and dynamic type assignment and checking within the GIPSY type system with regards to procedural functions called by other parts of the GIPSY program, e.g. the Lucid code segments. 

\item \codesegment{typedecl}
segment lists all user-defined data types that can potentially be used by the intensional part; usually objects. These are the types that do not explicitly appear in the matching table in \xt{tab:datatypes} describing the basic data types allowed in GIPSY programs.

\item \codesegment{$<$IMPERATIVELANG$>$}
segment declares that this is a code segment written in whatever IMPERATIVELANG may be, for example \codesegment{JAVA} for {\java}, \codesegment{CPP} for {\cpp}, \codesegment{FORTRAN} for {\fortran}, \codesegment{PERL} for {\perl}, \codesegment{PYTHON} for {\python}, etc.

\item \codesegment{$<$INTENSIONALLANG$>$}
segment declares that this is a code segment written in whatever INTENSIONALLANG may be,
for example
\codesegment{GIPL},
\codesegment{LUCX},
\codesegment{JOOIP},
\codesegment{INDEXICALLUCID},
\codesegment{JLUCID},
\codesegment{OBJECTIVELUCID},
\codesegment{TENSORLUCID},
\codesegment{ONYX}~\cite{grogonoonyx2004},
\codesegment{FORENSICLUCID}~\cite{flucid-isabelle-techrep-tphols08}, and
\codesegment{TRANSLUCID}, etc. as specified by the available {\gipsy} implementations and stubs. An example of a hybrid program is presented in \xl{list:language-mix}. The preamble of the program with the type and function declaration segments are the main source of type information that is used at compile time to annotate the nodes in the tree to help both static and semantic analysis.
\end{itemize}

\begin{lstlisting}[
    label={list:language-mix},
    caption={Example of a hybrid GIPSY program.},
    style=codeStyle
    ]
/**
 * Language-mix GIPSY program.
 * @author Serguei Mokhov
 */
#typedecl

myclass;

#funcdecl

myclass foo(int,double);
float bar(int,int):"ftp://newton.cs.concordia.ca/cool.class":baz;
int f1();

#JAVA
myclass foo(int a, double b)
{
   return new myclass(new Integer((int)(b + a)));
}

class myclass
{
   public myclass(Integer a)
   {
      System.out.println(a);
   }
}

#CPP
#include <iostream>

int f1(void)
{
   cout << "hello";
   return 0;
}

#OBJECTIVELUCID

A + bar(B, C)
where
   A = foo(B, C).intValue();
   B = f1();
   C = 2.0;
end;

/*
 * in theory we could write more than one intensional chunk,
 * then those chunks would evaluate as separate possibly
 * totally independent expressions in parallel that happened
 * to use the same set of imperative functions.
 */

// EOF
\end{lstlisting}

\subsubsection{Introduction to the GIPSY Type System}
\label{sect:gipsy-types}
\index{GIPSY Type System}
\index{GIPSY!Types}
\index{Types}
\index{Frameworks!GIPSY Type System}

The introduction of {\jlucid}, {\olucid}, and
{\gicf}~\cite{mokhovmcthesis05,mokhovolucid2005,mokhovgicf2005,mokhovjlucid2005}
prompted the development of the GIPSY Type System\index{GIPSY!Type System} as
implicitly understood by the {\lucid} language and its incarnation within
the {\gipsy} to handle types in a more general manner as a glue between
the imperative and intensional languages within the system.
Further evolution of {\lucx} introducing contexts as first-class values
and {\jooip} highlighted the need of the further development of the
type system to accommodate the more general properties of the intensional
and hybrid languages.

\paragraph{Matching Lucid and Java Data Types}
\label{sect:datatypes-matching}
\index{data types!matching Lucid and Java}

Here we present a case of interaction between {\lucid} and {\java}.
Allowing {\lucid} to call Java methods brings a set of issues related to
the data types, especially when it comes to type checks between Lucid and Java
parts of a hybrid program. This is pertinent when Lucid
variables or expressions are used as parameters to Java methods and when
a Java method returns a result to be assigned to a Lucid
variable or used in an intensional expression. The sets of types in both cases are not exactly
the same. The basic set of Lucid data types as defined by Grogono~\cite{gipcincrements}
is \api{int}, \api{bool}, \api{double}, \api{string},
and \api{dimension}. {\lucid}'s \api{int} is of the same size
as Java's \api{long}. GIPSY and {\java} \api{double}, \api{boolean}, and \api{String} are roughly the same.
Lucid \api{string} and Java \api{String} are simply mapped internally through \api{StringBuffer};
thus, one can think of the Lucid \api{string}
as a reference when evaluated in the intensional program. Based on this fact, the lengths
of a Lucid \api{string} and Java \api{String} are the same. Java \api{String} is also
an object in {\java}; however, at this point, a Lucid
program has no direct access to any \api{String}'s properties (though internally we
do and we may expose it later to the programmers).
We also distinguish the \api{float} data type for single-precision floating point operations.
The \api{dimension} index type is said to be an integer or string (as far as its dimension tag values
are concerned), but might be of other types eventually, as discussed in~\cite{gipsy-context-calculus-07}. 
Therefore, we perform data type matching as presented in~\xt{tab:datatypes}.
Additionally, we allow \api{void} Java return type
which will always be matched to a Boolean expression \api{true} in {\lucid} as
an expression has to always evaluate to something.
As for now our types mapping and restrictions are as per~\xt{tab:datatypes}. This is the mapping table for the Java-to-IPL-to-Java type adapter. Such a table would exist for mapping between any imperative-to-intensional language and back, e.g. the {\cpp}-to-IPL-to-{\cpp} type adapter.

\begin{table*}[htb!]
\caption{Matching data types between Lucid and Java.}
\begin{minipage}[b]{\textwidth}
\begin{center}
\begin{tabular}{|c|c|l|} \hline
{\scriptsize Return Types of Java Methods}  & {\scriptsize Types of Lucid Expressions} & {\scriptsize Internal GIPSY Types}\\ \hline\hline
\api{int}, \api{byte}, \api{long}           & \api{int}, \api{dimension}               & \api{GIPSYInteger} \\
\api{float}                                 & \api{float}                              & \api{GIPSYFloat} \\
\api{double}                                & \api{double}                             & \api{GIPSYDouble} \\
\api{boolean}                               & \api{bool}                               & \api{GIPSYBoolean} \\
\api{char}                                  & \api{char}                               & \api{GIPSYCharacter} \\
\api{String}                                & \api{string}, \api{dimension}            & \api{GIPSYString} \\
\api{Method}                                & {\it function}                           & \api{GIPSYFunction} \\
\api{Method}                                & {\it operator}                           & \api{GIPSYOperator} \\
\api{[]}                                    & \api{[]}                                 & \api{GIPSYArray} \\
\api{Object}                                & {\it class}                              & \api{GIPSYObject} \\
\api{Object}                                & {\it URL}                                & \api{GIPSYEmbed} \\
\api{void}                                  & \api{bool::true}                         & \api{GIPSYVoid} \\ \hline\hline

{\scriptsize Parameter Types Used in Lucid} & {\scriptsize Corresponding Java Types}   & {\scriptsize Internal GIPSY Types} \\ \hline\hline
\api{string}                                & \api{String}                             & \api{GIPSYString}\\
\api{float}                                 & \api{float}                              & \api{GIPSYFloat}\\
\api{double}                                & \api{double}                             & \api{GIPSYDouble}\\
\api{int}                                   & \api{int}                                & \api{GIPSYInteger} \\
\api{dimension}                             & \api{int}, \api{String}                  & \api{Dimension} \\
\api{bool}                                  & \api{boolean}                            & \api{GIPSYBoolean}\\
{\it class}                                 & \api{Object}                             & \api{GIPSYObject}\\
{\it URL}                                   & \api{Object}                             & \api{GIPSYEmbed}\\
\api{[]}                                    & \api{[]}                                 & \api{GIPSYArray}\\
{\it operator}                              & \api{Method}                             & \api{GIPSYOperator}\\
{\it function}                              & \api{Method}                             & \api{GIPSYFunction}\\ \hline
\end{tabular}
\end{center}
\end{minipage}
\label{tab:datatypes}
\end{table*}

\paragraph{Overview of the Design and Implementation of the Type System.}

While the main language of {\gipsy}, {\lucid}, is polymorphic and does not have explicit types, co-existing with other languages necessitates definition of GIPSY types and their mapping to a particular language being embedded. \xf{fig:gipsy-types} presents the detailed design of the GIPSY Type System.

\begin{figure}
	\begin{centering}
	\includegraphics[angle=90,width=\textwidth,totalheight=.9\textheight]{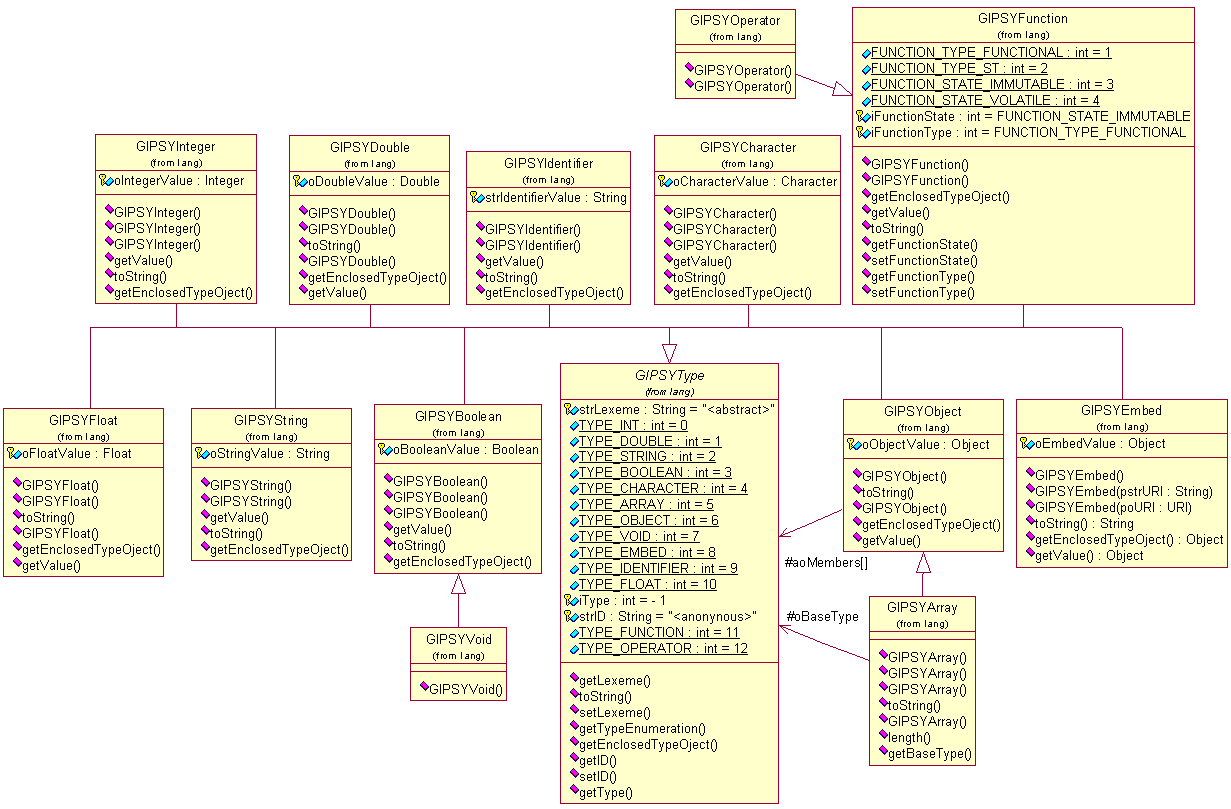}
	\caption{GIPSY Type System.}
	\label{fig:gipsy-types}
	\end{centering}
\end{figure}

Each class is prefixed with \texttt{GIPSY} to avoid possible confusion with similar definitions in the \api{java.lang} package. The \api{GIPSYVoid} type always evaluates to the Boolean \texttt{true}, as described earlier in \xs{sect:datatypes-matching}. The other types wrap around the corresponding {\java} object wrapper classes for the primitive types, such as \api{Long}, \api{Float}, etc. Every class keeps a lexeme (a lexical representation) of the corresponding type in a GIPSY program and overrides \api{toString()} to show the lexeme and the contained value. These types are extensively used by the \api{Preprocessor}, imperative and intensional (for constants) compilers, the \api{SequentialThreadGenerator}, and \api{SemanticAnalyzer} for the general type of GIPSY program processing, and by the GEE's \api{Executor}.

The other special types that have been created are either experimental
or do not correspond to a wrapper of a primitive type. \api{GIPSYIdentifier}
type case corresponds to a declaration of some sort of an identifier in a GIPSY
program to be put into the dictionary, be it a variable or a function name
with the reference to their definition.
Constants and conditionals may be anonymous and
thereby not have a corresponding identifier. \api{GIPSYEmbed} is another
special type that encapsulates embedded code via the URL parameter
and later is exploded into multiple types corresponding to procedural demands
(Java or any other language methods or functions)~\cite{mokhovmcthesis05,mokhovjlucid2005}.
\api{GIPSYFunction} and its descendant \api{GIPSYOperator} correspond to the
function types for regular operators and user-defined functions. A \api{GIPSYFunction}
can either encapsulate an ordinary Lucid function (which is immutable as in functional programming) or a procedure (e.g. a Java method), which may often
be mutable (i.e. with side effects). These four types ({\em identifier, embed, function}, and {\em operator}) are not directly exposed
to a GIPSY programmer and at this point are managed internally.
By the latter we mean we have not reached the stage when we can provide
them for explicit use by programmers; however, the semantics of is still
defined and specified at the requirements, design, and implementation levels.
\api{GIPSYContext} and
\api{Dimension} are a new addition to the type system implementation since~\cite{mokhovmcthesis05}.
They represent context-as-first-class-values in the context calculus defined by Wan in~\cite{wanphd06}
and refined and implemented by Tong~\cite{tongxinmcthesis08}.
The rest of the type system is exposed to the GIPSY programmer in the preamble
of a GIPSY program, i.e., the \codesegment{funcdecl} and \codesegment{typedecl}
segments, which result in the embryo of the dictionary for linking,  semantic analysis,
and execution. Once imperative compilers of procedural demands return, the type data
structures (return and parameter types) declared
in the preamble are matched against what was discovered by the compilers and
if the match is successful, the link is made.
By capturing the types such as {\em identifier, embed, function, operator}
and {\em context, dimension}, the GIPSY type system lays down fundamentals
the higher-order intensional logic (HOIL) support that combines 
functional programming, intensional logic, context calculus, and in
some instances hybrid paradigm support, and the corresponding types.
We describe various properties of the concrete GIPSY types and their
more detailed specification in \xa{appdx:gipsy-types-spec} and
\xa{appdx:gipsy-types-properties}. There we detail the inner workings
of each type in more detail as well describe some of the properties
through the notions of existential, union, intersection, and linear
types. They have been excluded from the main body of the article due
to size limitations.

\subsection{The Core Lucid Standard}
\label{sect:core-lucid-spec}

The Core Lucid standard specification, syntax, semantics, translation rules,
type system, and verifications are to be placed in this section upon consensus of
the SIGLUCID members.

{\todo}

\subsubsection{Syntax}
\label{sect:core-lucid-syntax}
{\todo}

\subsubsection{Semantics}
\label{sect:core-lucid-semantics}
{\todo}

\section{Conclusion}
\label{sect:conclusion}

We have layed out the first foundational notions of a practical Lucid
standard at the 1st SECASA in 2008 in Turku, Finland, associated with
COMPSAC 2008. Since then two (3) more SECASA's happened: in 2009 in Seattle,
2010 in Montreal, and another one is planned in 2011. Prior that we
are producing this first set of notes from the meeting and related work.


\section{Future Work}
\label{sect:future-work}

We plan on further
meet and refine these notes and the standards and further accrete the
related work. Our eventual goal after the standard draft is complete 
publish it along with a comprehensive survey of the recent related work
as well as historical review.


\section{Acknowledgments}

This research work was funded by the respective grants, faculties
and departments of the authors.
Acknowledgment also goes to the former and present colleagues
collaborated with us in the past.


%
\label{sect:bib}
\bibliography{lucid-standards-and-history}

%
%
\printindex

\end{document}